\RequirePackage{fix-cm}
\documentclass[smallextended]{svjour3}       
\smartqed  
\usepackage{graphicx}
\usepackage{mathptmx}      
\usepackage{latexsym}
\usepackage{amsmath}
\usepackage{amssymb}
\usepackage{enumitem}
\usepackage{hyperref}
\numberwithin{equation}{section}
\renewcommand{\le}{\leqslant}

\newcommand{\ws}{w_*} 
\newcommand{\wth}{w_\text{th}}
\newcommand{\wpl}{w_+}
\newcommand{\splt}{Q} 
\newcommand{\ssplt}{q} 

\newcommand{\phy}{p}      
\newcommand{\zoo}{z}      
\newcommand{\gp}{G_p}       
\newcommand{\gz}{G_z}       
\newcommand{\pred}{S}     
\newcommand{\spred}{s}     
\newcommand{\death}{M}  
\newcommand{\sdeath}{m} 
\newcommand{\dup}{K} 
\newcommand{\sdup}{k} 
\newcommand{\F}{\phi}     
\newcommand{\N}{N}        
\newcommand{\halfN}{r}    
\newcommand{\yield}{\theta}   
\newcommand{\E}{E}   
\newcommand{\sE}{e}   
\newcommand{\Hl}{H}   
\newcommand{\sHl}{h}   

\newcommand{\characteristic}{maximum }
\newcommand{\duplication}{division }
\newcommand{\resource}{nutrient }

\usepackage{todonotes}

\journalname{Journal of Mathematical Biology}
\begin{document}

\title{Sheldon Spectrum and the Plankton Paradox: Two Sides of the Same Coin
\thanks{Work funded by the Spanish
mobility grant PRX12/00124 and project FIS2015-64349-P (MINECO/FEDER, UE)
(JAC), and EU Grant 634495 -- MINOUW -- H2020-SFS-2014-2015 (GWD and RL).} }
\subtitle{A trait-based plankton size-spectrum model}

\titlerunning{Trait-based plankton size-spectrum model} 

\author{Jos\'e A.~Cuesta \and
        Gustav~W.~Delius \and~ Richard~Law
}


\institute{J. A. Cuesta \at
	      Grupo Interdisciplinar de Sistemas Complejos (GISC) and UC3M-BS
	      Institute of Financial Big Data (IFiBiD), Departamento de
              Matem\'aticas, Universidad Carlos III de Madrid, Madrid, Spain \\
              Tel.: +34-916248751, Fax: +34-916249129, 
              \email{cuesta@math.uc3m.es}
           \\ and
 	      Instituto de Biocomputaci\'on y F\'{\i}sica de Sistemas Complejos
              (BIFI), Universidad de Zaragoza, Zaragoza, Spain 
           \and
           G. W. Delius 
           \and
           R. Law \at
	      Department of Mathematics and York Centre for Complex Systems Analysis,
              University of York, York, United Kingdom.
}

\date{Received: date / Accepted: date}

\maketitle

\begin{abstract}
The Sheldon spectrum describes a remarkable regularity in aquatic ecosystems:
the biomass density as a function of logarithmic body mass is approximately
constant over many orders of magnitude. While size-spectrum models have 
explained this phenomenon for assemblages of multicellular organisms,
this paper introduces a species-resolved size-spectrum model to explain the
phenomenon in unicellular plankton. A Sheldon spectrum spanning the cell-size
range of unicellular plankton necessarily consists of a large number of coexisting
species covering a wide range of characteristic sizes. The coexistence of many phytoplankton species feeding on a small number of resources is known as the
Paradox of the Plankton.
Our model resolves the paradox by showing that coexistence is facilitated by
the allometric scaling of four physiological rates.
Two of the allometries have empirical support, the remaining two emerge from
predator-prey interactions exactly when the abundances follow a Sheldon
spectrum.
Our plankton model is a scale-invariant trait-based size-spectrum model: it describes the
abundance of phyto- and zooplankton cells as a function of both size and species
trait (the maximal size before cell division). It incorporates growth due to
resource consumption and predation on smaller cells, death due to predation, and
a flexible cell division process. We give analytic solutions at steady state for
both the within-species size distributions and the relative abundances across
species.

\end{abstract}

\section{Introduction}
\label{intro}
Gaining a better understanding of plankton dynamics is of great ecological
importance, both because plankton form an important component of the global
carbon cycle and couples to the global climate system and because 
plankton provide the base of the aquatic
food chain and therefore drives the productivity of our lakes and oceans.
In spite of enormous progress in plankton modelling,
there is still a lack of fundamental understanding of even some rather
striking phenomena. We address this in this paper with a novel
conceptual plankton model that for the first time gives analytical
results that simultaneously
describes both the within-species cell size distribution and the 
across-species distribution of plankton biomass.

One of the most remarkable patterns in ecology manifests itself in the
distribution of biomass as a function of body size in aquatic ecosystems
\cite{sheldon:1972a}. Very approximately, equal intervals of the logarithm of
body mass contain equal amounts of biomass per unit volume. This implies that
biomass density decreases approximately as the inverse of body mass. \emph{Size
spectra} with this approximate shape are observed over many orders of magnitude,
encompassing both unicellular and multicellular organisms
\cite{gaedke:1992,quinones:2003,san_martin:2006} and it has been conjectured
that this relationship applies all the way from bacteria to whales
\cite{sheldon:1972a}. Accordingly, aquatic environments are more populated by
small organisms than larger ones in a predictable way \cite{sheldon:1972b}.

Early theories, without dynamics, gave results consistent with this power law
\cite{platt:1977} and they were followed by dynamic theories for multicellular
organisms (size-spectrum models), where the biomass distribution is an outcome
of the processes and interactions between these organisms at different sizes
\cite{silvert:1978,silvert:1980,camacho:2001,benoit:2004,andersen:2006,%
capitan:2010,datta:2010,datta:2011,hartvig:2011}. In these models, multicellular
organisms grow by feeding on and killing smaller organisms, thereby coupling the
two opposing faces of predation: death of the prey, and body growth of the
predator ---during which survivors can grow over orders of magnitude. A common
feature of the models is the allometric scaling of the rates of the different
processes. For recent reviews of size-spectrum modelling see
\cite{sprules:2016,guiet:2016}.

Current models of size-spectrum dynamics are constructed with multicellular, 
heterotrophic organisms in mind, and make simplifying assumptions about 
the unicellular plankton on which they ultimately depend to provide a closure
for the models (e.g. \cite{hartvig:2011,datta:2010}).
The unicellular-multicellular distinction is important. 
Unicellular plankton encompass autotrophs (phytoplankton) that use 
inorganic nutrients and light to synthesize their own food, as well as
heterotrophs (zooplankton) that feed on other organisms, and mixotrophs
that do both. Also, unicellular organisms just double in size before splitting 
into two roughly equally-sized cells, rather than going through the prolonged 
somatic growth of multicellular organisms. Since cell masses of unicellular
plankton span an overall range of approximately $10^8$, the power law 
cannot therefore be generated without coexistence of many species.

Coexistence of species in the plankton is itself an unresolved problem.
In the case of phytoplankton, the problem is known as `the paradox of the 
plankton', because of the great diversity of phytoplankton taxa, 
seemingly unconstrained by the small number of resources they compete for 
\cite{hutchinson:1961}.  There is no consensus yet as to what mechanism(s)
can allow a large number of competing species to coexist on a small number of
resources \cite{roy:2007}.  Hutchinson thought environmental fluctuations could
be the answer, but this is currently acknowledged to be insufficient 
as an explanation \cite{fox:2013}. One promising proposal is a
strategy of ``killing the winner'' that involves a trade-off between
competitive ability and defence against enemies
\cite{thingstad:1997,winter:2010} and that resembles the mechanism of
predator-mediated coexistence  observed in ecology
\cite{leibold:1996,vage:2014}.  

In this paper we propose a dynamic trait-based size-spectrum model for
plankton that incorporates specific cellular mechanisms for growth, feeding,
and reproduction, along with their allometric laws, in order to capture the
size spectrum of biomass distribution in this size region of the aquatic
ecosystem (Section~\ref{sec:model}). We build on well established models of the
cell cycle \cite{fredrickson:1967,diekmann:1983,heijmans:1984,henson:2003,%
friedlander:2008,giometto:2013} but extend them to allow for many coexisting
species. The resulting model describes the dynamics of an ecosystem made of a
continuum of phytoplankton species living on a single resource, plus a
continuum of zooplankton species that feed on smaller cells. For the allometric
scaling of the growth and division rate we make use of recent experimental
measurements on phytoplankton production \cite{maranon:2013}.

The model is presented in two flavours: an idealised version
(Section~\ref{sec:ideal}) describing cells that grow until exactly doubling their
size and then split into two identical cells, and a more general model
(Section~\ref{sec:general}) in which cells are allowed to divide in a range of
sizes and produce two daughter cells of slightly different sizes. In both cases
we provide analytic expressions for the abundance distribution as a function of
size for any species.

For both flavours of the model we first study the conditions under which the
steady state allows for the coexistence of a continuum of infinitely many 
phytoplankton species and find ---not
surprisingly--- that a sufficient condition is a death rate that scales
allometrically as the growth rate. Then we introduce zooplankton that predate
on smaller cells (whether phyto- or zooplankton) and show that predation
produces the required scaling of the death rate if, and only if, the whole
plankton community conforms to Sheldon's power law size spectrum with an exponent very
close to the observed one. This power law size spectrum arises as the
steady state solution in our model (Section~\ref{sec:predation}). 

In other words, within the
model assumptions, coexistence of a continuum of plankton species implies
a specific allometric scaling of the death rate and the zooplankton growth
rate; the latter allometric scalings imply that the whole community distributes
as a power-law in size; and a power-law size distribution of the community
implies the coexistence of a continuum of plankton species. This is the main
result of our work. It reveals that the paradox of the plankton and the
observed size spectrum in aquatic ecosystems are but two manifestations of the
same phenomenon, and are both deeply rooted in the allometric scaling of basic
physiological rates. In Section~\ref{sec:invariance} we show that this 
allometric scaling makes the model invariant under scale
transformations, giving yet another explanation for the origin of the
Sheldon spectrum.

\section{Size- and species-resolved phytoplankton model}
\label{sec:model}

Our model for phytoplankton is a multispecies variant of the population balance
equation (PBE) model \cite{fredrickson:1967,henson:2003,friedlander:2008}.
Phytoplankton are assumed to be made mostly of unicellular autotrophs that grow
through the absorption of inorganic nutrients from the environment and
eventually split into two roughly equal-size daughter cells.

Cells will be described by their current size $w$ and by a size $\ws$
characteristic of the cell's species. For this characteristic size we choose
the maximum size a cell can reach. 

The two basic processes of the cellular dynamics are growth and division.  We
describe these in detail in the following subsections before using them in
section \ref{PBE} to give the dynamical population balance equation for
phytoplankton abundances.

\subsection{Cell growth}
\label{sec:growth}

A widely accepted model for organismal growth was proposed long ago by von
Bertalanffy \cite{vonbertalanffy:1957}. Although originally it was devised for
multicellular organisms, it has recently been argued that a similar model can
be used to describe the growth of microorganisms \cite{kempes:2011}.  According
to von Bertalanffy's model, the rate at which an organism grows is the result
of a competition between the gain of mass through \resource uptake and its loss
through metabolic consumption. Both terms exhibit allometric scaling, thus
\begin{equation}
\frac{dw}{dt}=Aw^{\alpha}-Bw^{\beta}.
\label{eq:growth}
\end{equation}
A typical assumption is $\alpha=2/3$ ---\resource uptake occurs through the
organismal membrane--- and $\beta=1$ ---metabolic consumption is proportional
to body mass \cite{kempes:2011}. However other choices are possible and
different values have been empirically obtained \cite{law:2016}. Whichever the
values, it
%
%
%
seems reasonable to constrain the exponents to satisfy $\alpha<\beta$
---leading to a slow-down of growth as cells get very large. Constants $A$ and
$B$ will vary from species to species, so depend on $\ws$.

With this model we can calculate the doubling period of a cell, defined as the
time $T(\ws)$ it takes to grow from $\ws/2$ to $\ws$:
\begin{equation}
T(\ws)=\int_{\ws/2}^{\ws}\frac{dw}{Aw^{\alpha}-Bw^{\beta}}
=\ws\int_{1/2}^1\frac{du}{A\ws^{\alpha}u^{\alpha}-B\ws^{\beta}u^{\beta}},
\label{eq:period}
\end{equation}
where $u=w/w_*$.

It turns out that this doubling period has been experimentally measured for
many different species of phytoplankton under the same environmental
conditions. All the results for phytoplankton cells larger than $\sim 5\,\mu$m
seem to scale with the same function $T=\tau\ws^{\xi}$, where $\tau$ is a
species-independent constant. Cells smaller than $\sim 5\mu$m have a doubling
period which increases, rather than decreases, as they become smaller
\cite{maranon:2013}. To all purposes then, our model will describe the
community spectrum from $\sim 5\,\mu$m upward. There is some controversy in the
experimental literature about the right value of the exponent $\xi$
\cite{law:2016}, but we need not be concerned by it.  When we need a concrete
value we will adopt the most recent value $\xi\approx 0.15$
\cite{maranon:2013}.

The allometric scaling observed for the duplication period can only be
compatible with Eq.~\eqref{eq:period} provided
\begin{equation}
A\equiv a\ws^{1-\alpha-\xi}, \qquad B\equiv b\ws^{1-\beta-\xi},
\end{equation}
where $a$ and $b$ do not depend on $\ws$. Then the proportionality constant
$\tau$ is given by
\begin{equation}
\tau=\int_{1/2}^1\frac{du}{au^{\alpha}-bu^{\beta}}.
\label{eq:c}
\end{equation}
Since $\tau$, $\alpha$, and $\beta$ can be experimentally determined, this
equation imposes a constraint on the constants $a$ and $b$.

In summary, joining a von Bertalanffy model for the growth rate with the
experimental observations for the \duplication rate yields the growth model
\begin{equation}
\frac{dw}{dt}=\gp(w,\ws)
=\ws^{1-\xi}\left[a\left(\frac{w}{\ws}\right)^{\alpha}
-b\left(\frac{w}{\ws}\right)^{\beta}\right].
\label{eq:gww0}
\end{equation}

It is worth noting that this growth rate is a homogeneous function satisfying
\begin{equation}
\gp(\lambda w,\lambda \ws)=\lambda^{1-\xi}\gp(w,\ws)
\label{eq:invarianceg}
\end{equation}
for any $\lambda>0$. Also notice that $a>b$ guarantees $\gp(w,\ws)>0$ for all
$0\le w\le\ws$.

\subsection{Cell division\label{division}}

Let $\dup(w,\ws)$ denote the division rate of a cell of current size $w$ and
\characteristic size $\ws$. We expect $\dup(w,\ws)$ to grow sharply near
$w=\ws$ ---to ensure that division is guaranteed to occur before a cell reaches
its \characteristic size. A widely studied cell \duplication mechanism assumes
a `sloppy size control' of the cell division cycle
\cite{powell:1964,tyson:1986}. Essentially, this means that cells can duplicate
at any moment after reaching a threshold size $\wth$ and before reaching their
largest possible size $w_*$.
By proposing a suitable function $\dup(w,\ws)$ 
Tyson and Diekmann \cite{tyson:1986} were able to fit the size distribution at
\duplication of a yeast.

While \cite{tyson:1986} assumed that duplication produces two equally-sized
daughter cells, we will in Section~\ref{sec:general} allow the size of the
daughter cells to be described by a density $\splt(w|w')$, the probability
density that a cell of size $w'$ splits into two cells of sizes $w$ and $w'-w$.
By construction $\splt(w|w')=0$ if $w\geqslant w'$ or $w\leqslant 0$, it bears
the symmetry $\splt(w'-w|w')=\splt(w|w')$ and satisfies the normalising
condition
\begin{equation}
\int_0^{\infty}\splt(w|w')\,dw=1
\label{eq:normalisation}
\end{equation}
for all $0<w'<\infty$. 

It is reasonable to assume that $\splt(w|w')$ is peaked around $w=w'/2$ ---
daughter cells will be roughly half the size of the parent cell. Another
reasonable assumption is that this distribution scales with cell size (i.e.,
fluctuations around the ideal splitting size $w=w'/2$ are relative to $w'$).
This amounts to assuming that $\splt(w|w')$ is a homogeneous function of $w$
and $w'$,
\begin{equation}
\splt(\lambda w,\lambda w')=\lambda^{-1}\splt(w,w').
\label{eq:invarianceq}
\end{equation}
The scaling exponent of $-1$ is due to the fact that $Q$ is a probability
density. We can therefore write $Q$ in the scaling form
\begin{equation}
\splt(w|w')=\frac{1}{w'}\,\ssplt\left(\frac{w}{w'}\right), \qquad
\text{ where }\int_0^{\infty}\ssplt(x)\,dx=1.
\end{equation}

\subsection{Cell population dynamics}
\label{PBE}

We will assume that the number of species and their population is large enough
so that we can make a continuum description through a density function
$p(w,\ws,t)$, such that $p(w,\ws,t)\,dwd\ws$ is the number of cells per unit
volume whose \characteristic sizes are between $\ws$ and $\ws+d\ws$ and whose
sizes at time $t$ are between $w$ and $w+dw$.

With these ingredients, the time evolution of the abundances $\phy(w,\ws,t)$
will be given by the population balance equation (PBE) 
\cite{fredrickson:1967,henson:2003,friedlander:2008}
\begin{equation}
\begin{split}
\frac{\partial}{\partial t}\phy(w,\ws,t) =&\,
-\frac{\partial}{\partial w}\big[\gp(w,\ws)\phy(w,\ws,t)\big] \\
&+2\int_0^{\ws}\splt(w|w')\dup(w',\ws)\phy(w',\ws,t)\,dw' \\
&-\dup(w,\ws)\phy(w,\ws,t)-\death(w,\ws)\phy(w,\ws,t).
\end{split}
\label{eq:MFdiv}
\end{equation}
The first two terms describe the dynamics of a growing organism as an extension
of the McKendrick--von Foerster equation \cite{silvert:1978,silvert:1980}. The
third term is the rate at which cells of size $w$ are produced from the
division of cells of size $0<w'<\ws$ ---the factor $2$ taking care of the fact
that each parent cell yields two daughter cells. The fourth term is the rate at
which cells of size $w$ divide. The last term is the rate at which cells of
size $w$ die for whatever reason. The same equation describes this process for
any species, hence $\ws$ enters as a parameter in every rate function involved.

\subsection{Nutrient dynamics}
\label{sec:resource}

The growth model just developed assumes an infinite abundance of nutrients. In
real aquatic ecosystems nutrients are limited though, and growth is hindered
when nutrients are scarce.  Accordingly, we need to modify our growth model in
order to take limited nutrients into account.

In the von Bertalanffy equation \eqref{eq:gww0} for the cell growth rate, the
first term describes the nutrient uptake through the cell membrane, and it is
modulated by the rate $a$. This rate will of course depend on the availability
of the nutrients needed for growth. Denoting by $\N$ the amount of nutrient per
unit volume, we need to replace $a$ by a function $a(\N)$. The simplest way to
do this is through the Monod equation
\cite{herbert:1956}
\begin{equation}
a(\N)=a_{\infty}\frac{\N}{\halfN+\N},
\label{eq:ar}
\end{equation}
with $\halfN$ the Michaelis-Mertens constant. This function has the important
property that the factor $a(\N)$ monotonically increases from $0$ toward its
saturation value $a_{\infty}$. However, other choices for $a(\N)$ with this
property are also possible.

Likewise, the details of how the nutrient dynamics is modelled are not
important for our conclusions. All we will require is that the uptake of
nutrient by the plankton leads to a corresponding depletion in the \resource
$\N$. Also, in order to sustain a non-zero plankton population, there needs to
be some replenishment of nutrient. The PBE model incorporates that through a
chemostat of maximum capacity $\N_0$
\cite{fredrickson:1967,heijmans:1984,henson:2003}:
\begin{equation}
\frac{d\N}{dt}=\varrho(\N)-\sigma(\N,\phy), \qquad
\varrho(\N)=\varrho_0\left(1-\frac{\N}{\N_0}\right).
\label{eq:resource}
\end{equation}
Here $\sigma(\N,\phy)$ represents the rate of nutrients consumption by all
phytoplankton cells, which is proportional to the uptake rate (the positive
term in the expression for $\gp(w,\ws,t)$ in Eq.~\eqref{eq:gww0}), integrated
over all species sizes $\ws$ and all cell sizes $w$:
\begin{equation}
\sigma(\N,\phy)=\frac{a(\N)}{\yield}\int_0^{\infty}d\ws\,\ws^{1-\alpha-\xi}
\int_0^{\ws}dw\,w^{\alpha}\phy(w,\ws,t).
\label{eq:uptake}
\end{equation}
The proportionality constant $\yield$ is the \emph{yield} constant, i.e. the
amount of biomass generated per unit of \resource.

\section{Idealised cell \duplication process}
\label{sec:ideal}

The important features of our model are insensitive to the details of the cell
division process. So it makes sense to first exhibit these features by solving
the model with the simplest idealised version of the cell division. Thus in
this section we assume that cells only split when they reach exactly the size
$\ws$, and they generate two identically sized daughter cells
\cite{diekmann:1983}. This prepares us well for a discussion of the general
case in Section~\ref{sec:general}.

\subsection{Dynamic equations}
\label{sec:idealeqs}

The idealised cell division amounts to choosing $\splt(w|w')=\delta(w-w'/2)$
---two identical daughter cells--- and $\dup(w,\ws)=\kappa(\ws)\delta(w-\ws)$
---\duplication occurs only when $w=\ws$. Here $\delta(x)$ denotes the Dirac
delta function. The parameter $\kappa(\ws)$ will be determined below. This
choice transforms the evolution equation~\eqref{eq:MFdiv} into
\begin{equation}
\begin{split}
\frac{\partial}{\partial t}\phy(w,\ws,t) =&\, 
-\frac{\partial}{\partial w}\big[\gp(w,\ws)\phy(w,\ws,t)\big] \\
&+\kappa(\ws)\phy(\ws,\ws,t)[2\delta(w-\ws/2)-\delta(w-\ws)]  \\
&-\death(w,\ws)\phy(w,\ws,t),
\end{split}
\label{eq:MF2equal}
\end{equation}
and of course $\phy(w,\ws,t)=0$ for $w>\ws$ and $w<\ws/2$.

The two delta functions on the right-hand side of Eq.~\eqref{eq:MF2equal} imply
that $\phy(w,\ws,t)$ will be continuous and differentiable everywhere except at
$w=\ws/2$ and $w=\ws$, where it will have two jump discontinuities such that
\begin{align}
\gp(\ws,\ws)\phy(\ws,\ws,t) &=\kappa(\ws)\phy(\ws,\ws,t),
\label{eq:jump1} \\
\gp(\ws/2,\ws)\phy(\ws/2,\ws,t) &=2\kappa(\ws)\phy(\ws,\ws,t).
\label{eq:jump2}
\end{align}
Equation~\eqref{eq:jump1} determines $\kappa(\ws)=\gp(\ws,\ws)$, so that
Eq.~\eqref{eq:jump2} implies the boundary condition
\begin{equation}
\gp(\ws/2,\ws)\phy(\ws/2,\ws,t)=2\gp(\ws,\ws)\phy(\ws,\ws,t).
\label{eq:bcfinal}
\end{equation}

Notice that, since $\delta(\lambda w-\lambda\ws)=\lambda^{-1}\delta(w-\ws)$,
this link between the \duplication rate function $\dup(w,\ws)$ and the growth
rate $\gp(w,\ws)$ renders the former homogeneous in its arguments,
\begin{equation}
\dup(\lambda w,\lambda \ws)
=\lambda^{-\xi}\dup(w,\ws).
\label{eq:invariancek}
\end{equation}

In summary, when considering the idealised \duplication process, the
phytoplankton density $\phi(w,\ws,t)$ is described by the equation
\begin{equation}
\frac{\partial}{\partial t}\phy(w,\ws,t)+
\frac{\partial}{\partial w}\big[\gp(w,\ws)\phy(w,\ws,t)\big]+
\death(w,\ws)\phy(w,\ws,t)=0,
\label{eq:MFideal}
\end{equation}
in the interval $\ws/2\le w\le\ws$, with the boundary condition
\eqref{eq:bcfinal}. This is coupled to Eqs.~\eqref{eq:resource} and
\eqref{eq:uptake} for the \resource.

\subsection{Steady state}
\label{sec:steadyideal}

We can look for solutions of Eq.~\eqref{eq:MFideal} that do not depend on time
by solving the first order ordinary differential equation
\begin{equation}
\frac{\partial}{\partial w}\big[\gp(w,\ws)\phy(w,\ws)\big]+
\death(w,\ws)\phy(w,\ws)=0, \qquad \frac{\ws}{2}\le w\le\ws,
\label{eq:sse}
\end{equation}
with the boundary condition \eqref{eq:bcfinal}. A straightforward integration
of Eq.~\eqref{eq:sse} yields
\begin{equation}
\phy(w,\ws)=\phy(\ws,\ws)\frac{\gp(\ws,\ws)}{\gp(w,\ws)}\exp\left\{
\int_{w}^{\ws}\frac{\death(w',\ws)}{\gp(w',\ws)}\,dw'\right\},
\label{eq:steadysol}
\end{equation}
where $\phy(\ws,\ws)$ is some (as yet) arbitrary value. If we now impose the
boundary condition ~\eqref{eq:bcfinal} on the solution~\eqref{eq:steadysol} we
arrive at the condition
\begin{equation}
\int_{\ws/2}^{\ws}\frac{\death(w',\ws)}{\gp(w',\ws)}\,dw'=\log 2.
\label{eq:coexistence-cond}
\end{equation}
The left-hand side of this condition is in general a function of $\ws$. This
means that only those species whose \characteristic sizes are such that
Eq.~\eqref{eq:coexistence-cond} holds can have a non-zero stationary abundance.
The only possibility for the remaining species is $\phy(\ws,\ws)=0$, i.e.,
extinction.

There is, however, one case in which Eq.~\eqref{eq:coexistence-cond} can hold
for \emph{all} species, namely when the death rate is a homogeneous function
$\death(\lambda w,\lambda\ws)=\lambda^{-\xi}\death(w,\ws)$, or, equivalently,
if it has the shape
\begin{equation}
\death(w,\ws)=\ws^{-\xi}\sdeath(w/\ws)
\label{eq:mu}
\end{equation}
for some function $\sdeath(x)$. Provided this condition is met, the solution
\eqref{eq:steadysol} takes the explicit form
\begin{equation}
\label{pp}
\phy(w,\ws)=\phy(\ws,\ws)\F(w/\ws),
\end{equation}
with
\begin{equation}
\F(x)=
\frac{a(\N)-b}{a(\N)x^{\alpha}-bx^{\beta}}
\exp\left\{\int_x^1\frac{\sdeath(y)}{a(\N)y^{\alpha}-b\,y^{\beta}}\,dy\right\}.
\label{eq:phystfinal}
\end{equation}
In other words, all species show the same size distribution up to a constant
$\phy(\ws,\ws)$ that determines the overall abundance of that species.

In this case the boundary condition \eqref{eq:coexistence-cond} becomes
\begin{equation}
\int_{1/2}^{1}\frac{\sdeath(x)}{a(\N)x^{\alpha}-bx^{\beta}}\,dx=\log 2.
\label{eq:imug2}
\end{equation}
This equation holds for one and only one value of $\N$ (remember that $a(\N)$
is an increasing function of $\N$ and $a(0)=0$ and $a_{\infty}>b$). For any
value other than this, no steady state solution is possible except full
extinction. On the other hand, for this specific $\N$ all species
\emph{coexist} in the steady state.

According to Eq.~\eqref{eq:resource}, the condition for $\N$ to be the
\resource level at the steady state is $\varrho(\N)=\sigma(\N,\phy)$. Using the
expression \eqref{eq:phystfinal} for the steady-state $\phy(w,\ws)$, this can
be expressed as the following constraint on the overall abundances:
\begin{equation}
\int_0^{\infty}\ws^{2-\xi}\phy(\ws,\ws)\,d\ws=
\frac{\yield\varrho(\N)}{a(\N)}
\left(\int_0^1x^{\alpha}\F(x)\,dx\right)^{-1}.
\label{eq:manifold}
\end{equation}
This is only a single linear constraint on the function $\phy(\ws,\ws)$ and
thus is far from determining it uniquely.

To summarise this section: if the death rate scales allometrically with size
and all phytoplankton species share a common limited resource then there is a
steady state of the system in which all species coexist on this single
resource. The resource level is tuned by consumption. In its turn, its value
imposes a global constraint on the abundances of phytoplankton species.

This result is a manifestation of the `paradox of the plankton'
\cite{hutchinson:1961}, and reveals a mechanism by which it might come about: a
similar allometric scaling for both the growth and the death rate. As of now,
it is hard to think of a reason why this similar scaling should occur, but we
will return to this point in Section~\ref{sec:predation} where we will show that
predation is one possible mechanism.

\section{General \duplication process}
\label{sec:general}

Although the idealised \duplication process described in the previous section
is a simple setup that provides important insights on the system behaviour, it
has some undesirable features that call for improvements. Perhaps the worst of
them is the fact that any irregularity of the initial distribution of cell
sizes will remain there forever because there is nothing that smooths it out.
Consequently, the distribution could never evolve towards the steady-state
distribution. Two mechanisms can achieve the necessary size mixing to provide
this smoothing: first, the fact that cells do not split only when they exactly
reach the size $\ws$, and second, the fact that the sizes of the two daughter
cells are not identical. Both of them require introducing functions
$\dup(w,\ws)$ and $\splt(w|w')$ more general than Dirac's deltas.

\subsection{Model constraints}

The problem boils down to solving the PBE \eqref{eq:MFdiv}. Although linear,
this is a difficult integro-differential problem whose general solution can
only be obtained in the form of an infinite functional series
\cite{heijmans:1984}. This notwithstanding, there is a general class of
functions $\dup(w,\ws)$ and $\splt(w|w')$ for which a closed form solution is
possible, and the constraints that define this class are general enough to
describe real situations. Let us spell out these constraints.

To guarantee that all cells divide before growing beyond size $\ws$ the rate
$\dup(w,\ws)$ needs to satisfy
\begin{equation}\label{kinf}
\int_0^{\ws}\dup(w,\ws)dw=\infty.
\end{equation}
There will be some smallest size $\wth$ below which cells can not divide. Hence
$\dup(w,\ws)$ is non-zero only for $\wth<w<\ws$. Let us also assume that
$\splt(w|w')$ is non-zero only for $(1-\delta)w'/2<w<(1+\delta)w'/2$ for some
$\delta$ that measures the maximum variability of the daughter cells' sizes
relative to the parent's. With these two assumptions it is clear that the
largest possible size of a daughter cell is $\wpl=(1+\delta)w_*/2$. We  further
assume $\wpl<\wth$.

Let us split the abundance into `large' and `small' cells according to
\begin{equation}
\phy(w,\ws,t)=
\begin{cases}
\phy_l(w,\ws,t), & w\geq\wpl, \\
\phy_s(w,\ws,t), & w\leq\wpl.
\end{cases}
\end{equation}
Then, the integral term in the right-hand side of Eq.~\eqref{eq:MFdiv} will
make no contribution for any $w>\wpl$, and we will have, for $\wpl\le w\le\ws$,
\begin{equation}
\begin{split}
\frac{\partial}{\partial t}\phy_l(w,\ws,t) =&\,
-\frac{\partial}{\partial w}\big[\gp(w,\ws)\phy_l(w,\ws,t)\big] \\
&-\dup(w,\ws)\phy_l(w,\ws,t)-\death(w,\ws)\phy_l(w,\ws,t).
\end{split}
\label{eq:MFdivlarge}
\end{equation}

Due to our assumption that $\wpl<\wth$ we can replace $\phy(w,\ws,t)$ by
$\phy_l(w,\ws,t)$ in the integral term of Eq.~\eqref{eq:MFdiv}; hence, for
$0\le w\le\wpl$,
\begin{equation}
\begin{split}
\frac{\partial}{\partial t}\phy_s(w,\ws,t) =&\,
-\frac{\partial}{\partial w}\big[\gp(w,\ws)\phy_s(w,\ws,t)\big] \\
&+2\int_{\wth}^{\ws}\splt(w|w')\dup(w',\ws)\phy_l(w',\ws,t)\,dw' \\
&-\death(w,\ws)\phy_s(w,\ws,t).
\end{split}
\label{eq:MFdivsmall}
\end{equation}

We have transformed the original problem into two, each in a different
interval. The first problem, Eq.~\eqref{eq:MFdivlarge}, is a homogeneous linear
differential equation decoupled from the second one, Eq.~\eqref{eq:MFdivsmall},
which turns out to be ---once the solution of the first problem is known--- a
non-homogeneous linear differential equation.

These two equations, \eqref{eq:MFdivlarge} and \eqref{eq:MFdivsmall}, have to
be supplemented with the boundary conditions 
\begin{equation}
\phy_s(0,\ws,t)=0,~~~~
\phy_s(\wpl,\ws,t)=\phy_l(\wpl,\ws,t),~~~~
\phy_l(\ws,\ws,t) =0.
\end{equation}

\subsection{Scaling behaviour of the \duplication rate}
\label{sec:scaling dup}

In the idealised model (Section~\ref{sec:idealeqs}), since $\dup(w,\ws)$ was
proportional to a Dirac's delta, we could obtain its scaling from that of
$\gp(w,\ws)$ straight away. Unfortunately, the argument is no longer valid for
this more general setup. There is a workaround though: we can prove that
$\dup(w,\ws)$ scales as in the idealised case from the empirical observation
that the population growth rate of a single species in a nutrient-rich
environment scales as $\Lambda\sim\ws^{-\xi}$ \cite{maranon:2013}.

Suppose we prepare a nutrient-rich culture of cells of \characteristic size
$\ws$. Equations~\eqref{eq:MFdivlarge} and \eqref{eq:MFdivsmall} will describe
the abundances at different sizes. In this situation, for some initial time
interval we can assume $\death(w,\ws)=0$, so the population will increase
exponentially at rate $\Lambda$. Introducing
$\phy_l(\wpl,\ws,t)=\phy_l(\wpl,\ws)e^{\Lambda t}$ and
$\phy_s(\wpl,\ws,t)=\phy_s(\wpl,\ws)e^{\Lambda t}$ into those equations we end
up with
\begin{align}
\frac{\partial}{\partial w}\big[\gp(w,\ws)\phy_l(w,\ws)\big] 
=&\, -\dup(w,\ws)\phy_l(w,\ws)-\Lambda\phy_l(w,\ws),
\label{eq:explarge} \\
\begin{split}
\frac{\partial}{\partial w}\big[\gp(w,\ws)\phy_s(w,\ws)\big] 
=&\, 2\int_{\wth}^{\ws}\splt(w|w')\dup(w',\ws)\phy_l(w',\ws)\,dw' \\
&-\Lambda\phy_s(w,\ws).
\end{split}
\label{eq:expsmall}
\end{align}
The solution of Eq.~\eqref{eq:explarge} is
\begin{align}
\phy_l(w,\ws) &=\phy_l(\wpl,\ws)\frac{\gp(\wpl,\ws)}{\gp(w,\ws)}\E(w,\ws),
\label{eq:sollarge} \\
\E(w,\ws) &=\exp\left\{-\int_{\wpl}^w\frac{\dup(w',\ws)+\Lambda}
{\gp(w',\ws)}\,dw'\right\},
\label{eq:defElambda}
\end{align}
with $\phy_l(\wpl,\ws)$ an undetermined constant.

As for Eq.~\eqref{eq:expsmall}, its solution is
\begin{align}
\phy_s(w,\ws) &=\phy_l(w,\ws)
\left[1-\int_w^{\wpl}\frac{\Hl(w',\ws)}{E(w',\ws)}\,dw'\right],
\label{eq:solsmall} \\
\Hl(w,\ws) &=2\int_{\wth}^{\ws}\splt(w|w')\frac{\dup(w',\ws)}{\gp(w',\ws)}
\E(w',\ws)\,dw'.
\label{eq:defHl}
\end{align}
The condition $\phy_l(\wpl,\ws,t)=\phy_s(\wpl,\ws,t)$ is already met, and the
boundary condition $\phy_l(\ws,\ws,t)=0$ follows from Eq.~\eqref{kinf}.  The
boundary condition $\phy_s(0,\ws,t)=0$ implies
\begin{equation}
\int_0^{\wpl}\frac{\Hl(w',\ws)}{E(w',\ws)}\,dw'=1.
\label{eq:eigenvalue}
\end{equation}
This equation determines the population growth rate $\Lambda$ and allows us to
rewrite Eq.~\eqref{eq:solsmall} as
\begin{equation}
\phy_s(w,\ws)=\phy_l(w,\ws)\int_0^w\frac{\Hl(w',\ws)}{E(w',\ws)}\,dw'.
\label{eq:solsmallclean}
\end{equation}

Equation~\eqref{eq:eigenvalue} is the key to infer the scaling of
$\dup(w,\ws)$. If, in agreement with empirical measurements,
$\Lambda=\ell\ws^{-\xi}$ with $\ell$ independent on $\ws$, then
Eq.~\eqref{eq:eigenvalue} becomes
\begin{equation*}
2\int_0^{\frac{1+\delta}{2}}dx\int_{\frac{w_{\text{th}}}{\ws}}^1\frac{dy}{y}\,
q\left(\frac{x}{y}\right)
\frac{\ws^{\xi}K(\ws y,\ws)}{a(N)y^{\alpha}-by^{\beta}}\exp\left\{\int_y^x
\frac{\ws^{\xi}K(\ws z,\ws)+\ell}{a(N)z^{\alpha}-bz^{\beta}}\,dz\right\}=1,
\end{equation*}
a condition that can only be met provided $w_{\text{th}}/\ws$  does not depend
on $\ws$ and
\begin{equation}
\dup(w,\ws)=\ws^{-\xi}\sdup(w/\ws),
\label{eq:dupscaling}
\end{equation}
in other words, if the scaling $\dup(\lambda w,\lambda\ws)=
\lambda^{-\xi}\dup(w,\ws)$ holds. Of course it is also intuitively clear that
the division rate has to scale as $\ws^{-\xi}$ given that the doubling period
$T(\ws)$ scales as $\ws^\xi$, as discussed in Section~\ref{sec:growth}.  Thus we
see that the same empirical observation that leads to the functional form
\eqref{eq:gww0} for $\gp(w,\ws)$ also leads to Eq.~\eqref{eq:dupscaling}.

\subsection{Steady state}
\label{sec:steadygeneric}

The steady state of Eqs.~\eqref{eq:MFdivlarge} and \eqref{eq:MFdivsmall} is
readily obtained by replacing $\Lambda$ with $\death(w,\ws)$ in
Eqs.~\eqref{eq:explarge} and \eqref{eq:expsmall}. The solution will be as given
by Eqs.~\eqref{eq:sollarge} and \eqref{eq:solsmallclean}, but with $\E(w,\ws)$
given by
\begin{equation}
\E(w,\ws)=\exp\left\{-\int_{\wpl}^w\frac{\dup(w',\ws)+\death(w',\ws)}
{\gp(w',\ws)}\,dw'\right\}.
\end{equation}
The boundary condition \eqref{eq:eigenvalue} now fixes the value of $a(\N)$ in
the function $\gp(w,\ws)$ and thereby determines the steady-state \resource
level $\N$. 

The same considerations as for the idealised case hold here.
Equation~\eqref{eq:eigenvalue} will, in general, depend on $\ws$ and therefore
hold for at most one or a few species. The other species are extinct in the
steady state. Given the scaling \eqref{eq:dupscaling} for the \duplication
rate, the requirement for coexistence of all species is the scaling
\eqref{eq:mu} of the death rate, because then $\E(w,\ws)=\sE(w/\ws)$ and
$\Hl(w,\ws)=\ws^{-1}h(w/\ws)$, where
\begin{align}
\sE(x) &=\exp\left\{-\int_{\frac{1+\delta}{2}}^x
\frac{\sdup(y)+\sdeath(y)}{a(N)y^{\alpha}-by^{\beta}}\,dy\right\}, \\[2mm]
\sHl(x) &=2\int_{\frac{\wth}{\ws}}^1\frac{\sdup(y)\sE(y)}{a(N)y^{\alpha}
-by^{\beta}}q\left(\frac{x}{y}\right)\frac{1}{y}\,dy,
\end{align}
and the boundary condition \eqref{eq:eigenvalue} becomes
\begin{equation}
\int_0^{\frac{1+\delta}{2}}\frac{\sHl(x)}{\sE(x)}\,dx=1
\label{eq:bcinvariant}
\end{equation}
regardless of the species.

Finally, the steady state abundances are given by
\begin{equation}
\phy(w,\ws)=\phy(\wpl,\ws)\psi(w/\ws),
\end{equation}
where  $\phy(\wpl,\ws)$ is an undetermined function of $\ws$ and
\begin{align}
\psi(x) &=\frac{a(\N)\left(\frac{1+\delta}{2}\right)^{\alpha}-
b\left(\frac{1+\delta}{2}\right)^{\beta}}{a(\N)x^{\alpha}-
bx^{\beta}}e(x)\Theta(x), \\[2mm]
\Theta(x) &=
\begin{cases}
1, & x>\frac{1+\delta}{2}, \\[2mm]
\displaystyle \int_0^x\frac{\sHl(y)}{\sE(y)}\,dy, & x<\frac{1+\delta}{2}.
\end{cases}
\end{align}

A few remarks will make clear what the abundance distribution looks like. To
begin with, property \eqref{kinf} of $\dup(w,\ws)$ implies that $\sE(1)=0$, so
$\phy(\ws,\ws)=0$. On the other hand, given that $q(x/y)=0$ except for
$(1-\delta)/2<x/y<(1+\delta)/2$ (i.e.  $2x/(1+\delta)<y<2x/(1-\delta)$),
function $h(x)=0$ except for $\wth(1-\delta)/2\ws<x<(1+\delta)/2$. This means
that $\phy(w,\ws)=0$ for all $w\le\wth(1-\delta)/2$ and that it is a
differentiable function in the whole interval $[0,\ws]$. From the fact that
$\partial\phy(w,\ws)/\partial w<0$ when $w>\wpl$ we can conclude that the
maximum of this function will occur at some point $w_{\text{max}}<\wpl$.

\begin{figure}
\begin{center}
\includegraphics[width=90mm]{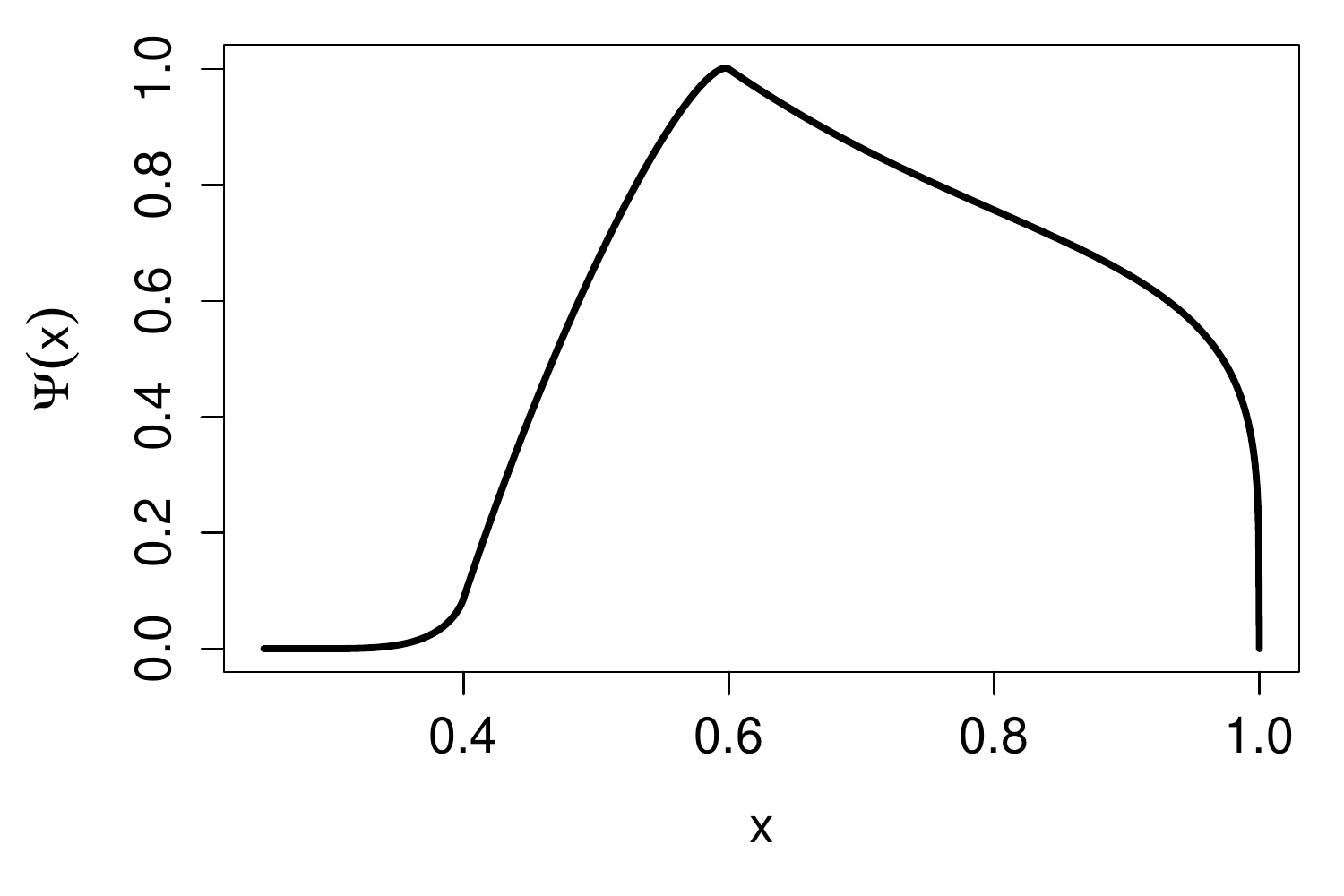}
\end{center}
\caption{The steady-state within-species size-distribution $\psi(x)$, with
constant mortality, growth parameter values $a=0.7, b=0.5, \alpha=0.85,
\beta=1$, a \duplication threshold of $0.7\ws$ and rate $\dup(w,\ws)$ given by
Eq.~\eqref{eq:dupscaling} with $k(x)=4(x-0.7)^2/(1-x)$ and daughter cell sizes
distributed uniformly between $0.4\ws$ and $0.6\ws$.}
\label{fig1}
\end{figure}

\section{Predation by zooplankton}
\label{sec:predation}

In the idealised model of cell \duplication of Section~\ref{sec:ideal} as well as
in the more general model of Section~\ref{sec:general}, we have seen that the
allometric scaling of the death rate is a crucial ingredient to the coexistence
of multiple phytoplankton species living on one or a few resources. The main
cause of phytoplankton death is predation. Many species feed on phytoplankton,
from unicellular organisms to whales. Even though a detailed model of the
marine ecosystem would have to include these very many types of grazers as well
as their predators, in order to keep the model simple ---and at the same time
to illustrate how predation can provide the sort of death rate necessary for
coexistence--- we will focus only on unicellular zooplankton.

We will denote the density of zooplankton cells by $\zoo(w,\ws,t)$, so that the
number of cells in a unit volume with a maximum size between $\ws$ and
$\ws+d\ws$ that at time $t$ have a size between $w$ and $w+dw$ is
$\zoo(w,\ws,t)dwd\ws$.

To model predation, we introduce a new rate function $\pred(w,w')$: the rate at
which a given predator cell of size $w$ preys on a given prey cell of size
$w'$. This rate could also be allowed to depend on the specific predator and
prey species through $\ws$ and $\ws'$. However, this would introduce an
unnecessary complication which would not add anything qualitatively different
to the discussion. 

A common ansatz for this rate function in the literature is
\begin{equation}
\pred(w,w')=w^{\nu}\spred(w/w').
\label{eq:predscaling}
\end{equation}
The second factor is a kernel that selects the preferred prey size relative to
the size of the predator \cite{wirtz:2012}. The power of $w$ in front of it
arises from the foraging strategy, which is known to depend allometrically on
cell size \cite{delong:2012}.

The mortality rate due to predation is obtained by integrating the
contributions from all predators.  For the sake of completeness, a background
death due to other source ---for which we will adopt the allometric scaling
\eqref{eq:mu}--- will be added to predation. Thus we set
\begin{equation}
\death(w,\ws,t)=\int_0^\infty\pred(w',w)\zoo_c(w',t)\,dw'+
\ws^{-\xi}\sdeath_b(w/\ws),
\label{eq:mupred}
\end{equation}
where the zooplankton community spectrum is defined as
\begin{equation}
\zoo_c(w,t)=\int_0^\infty \zoo(w,\ws,t)\,d\ws.
\label{eq:zct}
\end{equation}

Zooplankton abundance is described by an equation similar to
Eq.~\eqref{eq:MFdiv},
\begin{equation}
\begin{split}
\frac{\partial}{\partial t}\zoo(w,\ws,t) =&\,
-\frac{\partial}{\partial w}\big[\gz(w,\ws,t)\zoo(w,\ws,t)\big] \\
&+2\int_0^{\infty}\splt(w|w')\dup_\zoo(w',\ws,t)\zoo(w',\ws,t)\,dw' \\
&-\dup_\zoo(w,\ws,t)\zoo(w,\ws,t)-\death(w,\ws,t)\zoo(w,\ws,t),
\end{split}
\label{eq:pred}
\end{equation}
where the growth rate is now
\begin{equation}
\gz(w,\ws,t)=\int_0^\infty\pred(w,w')\epsilon w'
\left[\phy_c(w',t)+\zoo_c(w',t)\right]dw'
-b\ws^{1-\xi}\left(\frac{w}{\ws}\right)^{\beta},
\label{eq:h}
\end{equation}
with the phytoplankton community spectrum defined as
\begin{equation}
\phy_c(w,t)=\int_0^\infty \phy(w,\ws,t)\,d\ws.
\label{eq:pct}
\end{equation}
The first term in \eqref{eq:h} represents the uptake of nutrients from
predation. The factor $\epsilon$ expresses the efficiency with which prey
biomass $w'$ is converted into predator biomass. It is assumed that predators
prey indiscriminately on all species of cells, whether zoo- or phytoplankton.
The second term accounts for the metabolic consumption. Although we choose this
to be the same as for phytoplankton cells (see Eq.~\eqref{eq:gww0}),
substituting different values for $b$ and $\beta$ would not change the results
of the model qualitatively.

The steady state of the model we have just introduced has an important property
that is the main result of this paper, namely that, under the assumptions of
the model ---in particular the allometric scalings assumed for for the
phytoplankton growth rate (Eq.~\eqref{eq:invarianceg}) as well as for the
predation kernel (Eq.~\eqref{eq:predscaling})---, the death rate
$\death(w,\ws)$ and the zooplankton growth rate $\gz(w,\ws)$ scale
allometrically as 
\begin{equation}
\death(\lambda w,\lambda\ws)=\lambda^{-\xi}\death(w,\ws)
~~\text{ and }~~\gz(\lambda w,\lambda\ws)=\lambda^{1-\xi}\gz(w,\ws)
\end{equation}
if, and only if, the community spectra of the phyto- and zooplankton scale as
\begin{equation}
\phy_c(\lambda w)
=\lambda^{-\gamma}\phy_c(w)~~\text{ and }~~
\zoo_c(\lambda w)=\lambda^{-\gamma}\zoo_c(w),
\end{equation}
with $\gamma=1+\nu+\xi$.

The importance of this result lies in the fact that, according to the
discussion of Secs.~\ref{sec:steadyideal} and \ref{sec:steadygeneric}, the
allometric scaling of $\death(w,\ws)$ is a necessary and sufficient condition
for the steady state to exhibit a biodiverse phytoplankton community, and
similarly, given the scaling of $\death(w,\ws)$, that of $\gz(w,\ws)$ becomes
then a necessary and sufficient condition for the steady state to exhibit a
biodiverse zooplankton community. Accordingly, the paradox of the plankton and
the power-law size spectrum of the plankton community are \emph{two
manifestations of one single phenomenon} ---which also expresses itself in the
allometric scaling of those two rates.

We will discuss this point further in the Discussion section, and devote the
rest of this section to proving this result. If we substitute
$\zoo_c=\zoo_0w^{-\gamma}$ within Eq.~\eqref{eq:mupred} we obtain
\begin{equation}
\label{eq:deathz}
\death(w,\ws)=\ws^{-\xi}\sdeath(w/\ws), \quad
\sdeath(x)=\sdeath_b(x)+\zoo_0\,x^{-\xi}
\int_0^{\infty}y^{-\xi-1}\spred(y)\,dy.
\end{equation}
This trivially satisfies the required allometric scaling. If we substitute both
$\phy_c=\phi_0w^{-\gamma}$ and $\zoo_c=\zoo_0w^{-\gamma}$ within \eqref{eq:h}
we arrive at
\begin{equation}
\begin{split}
\gz(w,\ws) &=\ws^{1-\xi}\left[a_{\phy\zoo}\left(\frac{w}{\ws}\right)^{1-\xi}
-b\left(\frac{w}{\ws}\right)^{\beta}\right], \\
a_{\phy\zoo} &=\epsilon(\phy_0+\zoo_0)\int_0^{\infty}
x^{\gamma-3}\spred(x)\,dx.
\end{split}
\label{eq:growthz}
\end{equation}
This also complies with the required allometric scaling.

To prove the converse we impose the scaling $M(\lambda w,\lambda\ws)=
\lambda^{-\xi}M(w,\ws)$ on Eq.~\eqref{eq:mupred}, which leads to
\begin{equation*}
\int_0^{\infty}S(w',\lambda w)\zoo_c(w')\,dw'=
\lambda^{-\xi}\int_0^{\infty}S(w',w)\zoo_c(w')\,dw'.
\end{equation*}
Changing the variable $w'=\lambda u$ and using the scaling $S(\lambda w,
\lambda w')=\lambda^{\nu}S(w,w')$ derived from \eqref{eq:predscaling}, this
equation transforms into
\begin{equation*}
\lambda^{1+\nu}\int_0^{\infty}S(u,w)\zoo_c(\lambda u)\,du=
\lambda^{-\xi}\int_0^{\infty}S(w',w)\zoo_c(w')\,dw',
\end{equation*}
which holds if, and only if, $\zoo_c(\lambda w)=\lambda^{-\gamma}\zoo_c(w)$
with $\gamma=1+\nu+\xi$. Doing the same with the zooplankton growth rate
\eqref{eq:h} amounts to imposing the scaling
\begin{equation*}
\int_0^{\infty}S(\lambda w,w')w'\big[\phy_c(w')+\zoo_c(w')\big]\,dw'=
\lambda^{1-\xi}\int_0^{\infty}S(w,w')w'\big[\phy_c(w')+\zoo_c(w')\big]\,dw',
\end{equation*}
which, using the same argument as above, leads to
$\phy_c(\lambda w)=\lambda^{-\gamma}\phy_c(w)$.

An interesting by-product of this result is that the expressions for
$\death(w,\ws)$ and $\gz(w,\ws)$ have the same functional form as those
introduced in the analysis of phytoplankton in previous sections. Therefore we
can obtain the steady state of the full system doing similar calculations. We
will discuss this steady state first as obtained under the idealised
\duplication assumption and then as obtained for the general model.

\subsection{Steady state with idealised \duplication process}
\label{sec:idealzoo}

We can again make the idealised \duplication assumption that cells divide
exactly at size $\ws$ into two equal-size cells. As in the case of
phytoplankton, this amounts to choosing
$\dup_\zoo(w,\ws,t)=\gz(w,\ws,t)\delta(w-\ws)$ and
$\splt(w|w')=\delta(w-w'/2)$, which transforms the population balance equation
\eqref{eq:MFdiv} into
\begin{equation}
\frac{\partial}{\partial t}\zoo(w,\ws,t) =
-\frac{\partial}{\partial w}\big[\gz(w,\ws,t)\zoo(w,\ws,t)\big]
-\death(w,\ws,t)\zoo(w,\ws,t),
\label{eq:fulleompsideal}
\end{equation}
valid in the interval $\ws/2\le w\le\ws$, with the boundary condition
\begin{equation}
2\gz(\ws,\ws,t)\zoo(\ws,\ws,t)=\gz(\ws/2,\ws,t)\zoo(\ws/2,\ws,t).
\label{eq:bczoo}
\end{equation}

The expressions for the death and growth rates for zooplankton are formally the
same as those for phytoplankton. Therefore the steady state size distributions
of species abundances are given by
\begin{equation}
\phy(w,\ws)=\phy(\ws,\ws)\F_{\phy}(w/\ws), \qquad
\zoo(w,\ws)=\zoo(\ws,\ws)\F_{\zoo}(w/\ws), 
\end{equation}
where
\begin{equation}
\F_{\phy}(x)=\frac{a(\N)-b}{a(\N)x^{\alpha}-bx^{\beta}}
\exp\left\{\int_x^1\frac{\sdeath(y)}{a(\N)y^{\alpha}-by^{\beta}}\,dy\right\},
\end{equation}
$\N$ being the steady state value of the \resource concentration, and
\begin{equation}
\F_{\zoo}(x)=\frac{a_{\phy\zoo}-b}{a_{\phy\zoo}x^{1-\xi}-bx^{\beta}}
\exp\left\{\int_x^1\frac{\sdeath(y)}{a_{\phy\zoo}y^{1-\xi}-
by^{\beta}}\,dy\right\}
\end{equation}
with $a_{\phy\zoo}$ given in Eq.~\eqref{eq:growthz}.

The overall species abundances $\phy(\ws,\ws)$ and $\zoo(\ws,\ws)$ can be
obtained through Eqs.~\eqref{eq:zct} and \eqref{eq:pct}. For the phytoplankton,
for instance, given that $\phy(w,\ws)=0$ for $w>\ws$,
\begin{equation*}
\phy_c(w)=\int_w^{\infty}\phy(\ws,\ws)\F_{\phy}(w/\ws)\,dx=
w\int_0^1\phy\left(\frac{w}{x},\frac{w}{x}\right)
\F_{\phy}(x)\,\frac{dx}{x^2}.
\end{equation*}
Now, given the scaling $\phy_c(\lambda w)=\lambda^{-\gamma}\phy_c(w)$, this
equation implies that $\phy(\lambda\ws,\lambda\ws)=\lambda^{-\gamma-1}
\phy(\ws,\ws)$, i.e.,
\begin{equation}
\phy(\ws,\ws)=\frac{\phy_0}{I_{\phy}(\gamma-1)}\ws^{-\gamma-1},
\end{equation}
in terms of the functions
\begin{equation}
I_{\phy}(\eta)=\int_0^1x^{\eta}\F_{\phy}(x)\,dx, \qquad
I_{\zoo}(\eta)=\int_0^1x^{\eta}\F_{\zoo}(x)\,dx.
\end{equation}
A similar argument yields
\begin{equation}
\zoo(\ws,\ws)=\frac{\zoo_0}{I_{\zoo}(\gamma-1)}\ws^{-\gamma-1}.
\end{equation}

As in the case of phytoplankton alone, the level of \resource at the steady
state is determined by the boundary condition \eqref{eq:imug2}, which fixes the
value of $a(\N)$. There is a problem though. In this idealised version of a
plankton community we are implicitly assuming an infinite biomass, because we
are not imposing any lower nor upper limit on the size of cells. This
translates into an infinite \resource uptake by the phytoplankton,
\begin{align*}
\sigma(\N,\phy) &=\frac{a(\N)}{\yield}\int_0^{\infty}d\ws\,\ws^{1-\alpha-\xi}
\int_0^{\ws}dw\,w^{\alpha}\phy(w,\ws) \\
&=\frac{a(\N)}{\yield}\phy_0\frac{I_{\phy}(\alpha)}{I_{\phy}(\gamma-1)}
\int_0^{\infty}d\ws\,\ws^{1-\xi-\gamma},
\end{align*}
which will then require an infinite amount of \resource to survive.

In reality there will always be a minimum size $w_{\min}$ and a maximum size
$w_{max}$, so if we introduce the factor
\begin{equation}
\Xi=\int_{w_{\min}}^{w_{\max}}d\ws\,\ws^{1-\xi-\gamma}
\end{equation}
and assume that all resource-related quantities diverge proportional to $\Xi$,
we can rescale those quantities accordingly, so that they stay finite also in
the limit of $w_{\min}\to 0$ and $w_{\max}\to\infty$. Hence we introduce a
renormalised \resource concentration $\hat\N=\lim\N/\Xi$, where the limit takes
$w_{\min}\to 0$ and $w_{\max}\to\infty$, and similarly with other variables (a
hat will henceforth denote these renormalised quantities). The dynamics of the
\resource \eqref{eq:resource}, in terms of renormalised quantities,
becomes\footnote{In these expressions $\hat a(\hat\N)=a_{\infty}\hat\N/(\hat
r+\hat\N)$ and $\hat\varrho(\hat\N)=\hat\varrho_0(1-\hat\N/\hat\N_0)$.}
\begin{equation}
\frac{d\hat\N}{dt}=\hat\varrho(\hat\N)-\hat\sigma(\hat\N,\phy),
\label{eq:resourcehat}
\end{equation}
Hence in the steady state the renormalised \resource concentration satisfies
$\hat\varrho(\hat\N)-\hat\sigma(\hat\N,\phy)$, which can be rewritten as
\begin{equation}
\phy_0=\frac{\yield\hat\varrho(\hat\N)I_{\phy}(\gamma-1)}{\hat a(\hat\N)
I_{\phy}(\alpha)}.
\end{equation}
Once we have determined $\phy_0$, the boundary condition
\begin{equation}
\int_{1/2}^1\frac{\sdeath(y)}{a_{\phy\zoo}y^{1-\xi}-by^{\beta}}\,dy=\log 2
\end{equation}
yields $a_{\phy\zoo}$, which in turns determines $\zoo_0$ via
Eq.~\eqref{eq:growthz}.

\subsection{Steady state with general \duplication process}

We can introduce a \duplication rate for zooplankton $\dup_{\zoo}(w,\ws)$ with
similar properties as that for phytoplankton. The simplest choice is to take
the same function ---as it is conceivable that the dynamics of cell
\duplication does not depend on the feeding mechanism--- or any other
alternative, but in any case scaling \eqref{eq:dupscaling} must hold for
$\dup_{\zoo}(w,\ws)$ as well. Also, we assume that the size distribution of
daugher cells is described by the same function $\splt(w|w')$.

Then we can introduce a similar splitting for zooplankton abundance
\begin{equation}
\zoo(w,\ws,t)=
\begin{cases}
\zoo_l(w,\ws,t), & w\geq\wpl, \\
\zoo_s(w,\ws,t), & w\leq\wpl,
\end{cases}
\end{equation}
and write equations similar to \eqref{eq:MFdivlarge} and \eqref{eq:MFdivsmall}.
The steady state of those equations will be given by
\begin{equation}
\phy(w,\ws)=\phy(\wpl,\ws)\psi_{\phy}(w/\ws), \qquad
\zoo(w,\ws)=\zoo(\wpl,\ws)\psi_{\zoo}(w/\ws),
\end{equation}
where
\begin{align}
\psi_{\phy}(x) &=\frac{a(\N)\left(\frac{1+\delta}{2}\right)^{\alpha}-
b\left(\frac{1+\delta}{2}\right)^{\beta}}{a(\N)x^{\alpha}-bx^{\beta}}
\sE_{\phy}(x)\Theta_{\phy}(x), \\[2mm]
\sE_{\phy}(x) &=\exp\left\{-\int_{\frac{1+\delta}{2}}^x\frac{\sdup(y)
+\sdeath(y)}{a(\N)y^{\alpha}-by^{\beta}}\,dy\right\}, \\[2mm]
\sHl_{\phy}(x) &= \int_{\frac{\wth}{\ws}}^1\frac{\sdup(y)\sE_{\phy}(y)}
{a(\N)y^{\alpha}-by^{\beta}}\ssplt\left(\frac{x}{y}\right)\,dy, \\[2mm]
\Theta_{\phy}(x) &=
\begin{cases}
1, & x>\frac{1+\delta}{2}, \\[2mm]
\displaystyle \int_0^x\frac{\sdup(y)}{\sE_{\phy}(y)}\,dy, &
x<\frac{1+\delta}{2},
\end{cases}
\end{align}
and
\begin{align}
\psi_{\zoo}(x) &=\frac{a_{\phy\zoo}\left(\frac{1+\delta}{2}\right)^{1-\xi}-
b\left(\frac{1+\delta}{2}\right)^{\beta}}{a_{\phy\zoo}x^{1-\xi}-bx^{\beta}}
\sE_{\zoo}(x)\Theta_{\zoo}(x), \\[2mm]
\sE_{\zoo}(x) &=\exp\left\{-\int_{\frac{1+\delta}{2}}^x\frac{\sdup(y)
+\sdeath(y)}{a_{\phy\zoo}y^{1-\xi}-by^{\beta}}\,dy\right\}, \\[2mm]
\sHl_{\zoo}(x) &= \int_{\frac{\wth}{\ws}}^1\frac{\sdup(y)\sE_{\zoo}(y)}
{a_{\phy\zoo}y^{1-\xi}-by^{\beta}}\ssplt\left(\frac{x}{y}\right)\,dy, \\[2mm]
\Theta_{\zoo}(x) &=
\begin{cases}
1, & x>\frac{1+\delta}{2}, \\[2mm]
\displaystyle \int_0^x\frac{\sdup(y)}{\sE_{\zoo}(y)}\,dy, &
x<\frac{1+\delta}{2}.
\end{cases}
\end{align}

Introducing the functions
\begin{equation}
J_{\phy}(\eta)=\int_0^1x^{\eta}\psi_{\phy}(x)\,dx, \qquad
J_{\zoo}(\eta)=\int_0^1x^{\eta}\psi_{\zoo}(x)\,dx,
\end{equation}
and reproducing the arguments of Section~\ref{sec:idealzoo}, we obtain
\begin{equation}
\phy(\wpl,\ws)=\frac{\phy_0}{J_{\phy}(\gamma-1)}\ws^{-\gamma-1}, \qquad
\zoo(\wpl,\ws)=\frac{\zoo_0}{J_{\zoo}(\gamma-1)}\ws^{-\gamma-1},
\end{equation}
with
\begin{equation}
\phy_0=\frac{\yield\hat\varrho(\hat\N)J_{\phy}(\gamma-1)}{\hat a(\hat\N)
J_{\phy}(\alpha)}
\end{equation}
and $\zoo_0$ derived from \eqref{eq:growthz}, with $a_{\phy\zoo}$ obtained
through the boundary condition
\begin{equation}
\int_0^{\frac{1+\delta}{2}}\frac{\sHl_{\zoo}(y)}{\sE_{\zoo}(y)}\,
dy=1.
\end{equation}

What we can conclude from the analysis of the last two sections is that only
two steady states are possible in this system in which zooplankton predate on
phytoplankton: (a) a collapsed community in which at most a few species of
phytoplankton ---and possibly of zooplankton--- survive; or (b) a community
made of a continuum of species of sizes $0<w<\infty$ that align on a single
power law spectrum, with an exponent $\gamma$ determined by the allometry of
the phytoplankton growth rate and of the zooplankton predation rate. 

It is interesting to realise how the different facts assemble together to yield
this result. On the one hand, as zooplankton predation is the main cause of
phytoplankton mortality, in order for several phytoplankton species to coexist
the zooplankton community is forced to distribute their abundances on a power
law. In turn, zooplankton grow by predation, and in order for several
zooplankton species to coexist the phytoplankton community is forced to lie on
the same power law. We see then that both communities sustain each other, and
that biodiversity is both the cause and the consequence of the size spectrum.

\section{Scale invariance\label{sec:invariance}}

In the last paragraph of Section~\ref{sec:predation} we gave an intuitive
explanation of why both the phytoplankton and the zooplankton spectrum have to
follow a power law in the steady-state. There is also a more formal explanation
that we would like to exhibit in this section: the steady-state equations are
scale-invariant in the sense that if an abundance spectrum $p(w,\ws)$,
$z(w,\ws)$,
is a solution of the steady-state equations for some level of \resource
$\hat\N$, then so is the scale-transformed spectrum
\begin{equation}
\phy_\lambda(w,\ws)=\lambda^{\gamma+1}\phy(\lambda w,\lambda\ws),
\qquad
\zoo_\lambda(w,\ws)=\lambda^{\gamma+1}\zoo(\lambda w,\lambda\ws),
\label{sts}
\end{equation} 
for any positive $\lambda$. Thus solutions come in one-parameter families. The
steady-state however is expected to be unique, and this implies that it must be
scale invariant, which in turn implies that it must be of the power-law form
\begin{equation}
\phy(w,\ws)=\ws^{-\gamma-1}f_\phy(w/\ws)
~~\text{ and }~~
\zoo(w,\ws)=\ws^{-\gamma-1}f_\zoo(w/\ws)
\end{equation}
for some scaling functions $f_\phy$ and $f_\zoo$. These scaling functions were
calculated explicitly in earlier sections and depend on some details of the
model, but the power-law form of the abundances follows directly from the
scale-invariance of the model and is insensitive to other details.

This viewpoint, that the crucial property of the aquatic ecosystem is its scale
invariance was previously taken in \cite{capitan:2010}, where a scale-invariant
model for the fish part of the spectrum was presented. That paper did not model
the dynamics of the plankton part of the spectrum but simply assumed that it
was given by a power-law. The plankton model in this paper can be combined with
the fish model in \cite{capitan:2010} to give a dynamic scale-invariant model
of the entire spectrum. What remains to be done is to explain why evolution,
presented with the opportunity to fill a physical environment that itself
exhibits scale invariance over many orders of magnitude, like an ocean or a
large lake, would evolve organisms that preserve this scale invariance to a
great degree.

\section{Discussion and Conclusions}

Traditionally, size-based models for the population dynamics of unicellular 
organisms concentrate either on cell-level processes like cell
growth and cell division to describe the size distribution of cells within a
species \cite{fredrickson:1967,diekmann:1983,heijmans:1984,henson:2003,%
friedlander:2008}, or they concentrate on population-level processes like
predator-prey interactions to describe the abundance distribution across species
of different characteristic size\cite{moloney:1991,gin:1998,armstrong:1999,baird:2007,stock:2008,poulin:2010,banas:2011,ward:2014}. We have introduced a model that does both
simultaneously: it resolves the distribution of cell sizes within a species and
the distribution of biomass across species and thereby allows us to start from
individual-level processes and their allometric scaling and from them derive
population level phenomena like the power-law Sheldon spectrum. The only other
work of a similar nature that we are aware of is \cite{giometto:2013}.

At the cell level, our model combines a von Bertalanffy cell growth model with a
flexible cell division model. This cell division model allows a sloppy size
control, so that \duplication can occur for a wide range of sizes. 
In addition, the two daughter cells do not necessarily
have equal size but instead are described by a size distribution.
Even though this is quite a general model for cell growth and cell division, we
were able to give exact analytic solutions for the steady state cell
size distribution. This goes beyond what is
novel and may be useful also for studying size
distributions of cells other than plankton cells.
We also
worked with an idealised version of the cell cycle (cells split into two
identical daughter cells once they exactly double their size) in parallel to the
more realistic model to show that the main conclusions of our paper do not
depend on the details of the \duplication model.

The most important aspect of our model is the  coupling of the growth of a 
predator cell to the death of a smaller prey cell. This makes the cell 
growth rate depend on the
abundance of prey and the cell death rate depend on the abundance of 
predators, leading to a non-linear model. It is remarkable that, in spite of 
this non-linearity, this coupling together of cells of all species allows an 
exact steady-state solution giving the size distributions for a continuum of 
coexisting species.

The model has the property that, at steady state, the coexistence of multiple
unicellular plankton species and the Sheldon power-law size spectrum are two
different manifestations of just one single phenomenon ---`two sides of the same
coin'. This conclusion rests very much on the allometry of the four rates
involved: the growth rates of phyto- and zooplankton, the death rate, and the
predation kernel. The first and last of these allometries are supported by
empirical data and have specified allometric exponents. However, the allometries
of the zooplankton growth and death rates have to emerge from the predator-prey
interactions at steady state, and are technical outcomes of our modelling. In
summary, one can assume in the model any one of the following properties: (a)
allometry of the rates, (b) coexistence of multiple species, and (c) a power-law
community size spectrum. Then, from this, the other two properties can be
derived.

While we have been able to show that the model predicts a coexistence steady
state that agrees with the Sheldon spectrum, we have not discussed the stability
of this steady state against small perturbations. Our initial investigations
suggest that the steady state is unstable unless additional stabilising terms,
like for example a density dependence of the predation rate, are added. 
As those stabilising
terms are reduced, the system undergoes a Hopf bifurcation during which the
steady-state becomes unstable and the new attractor is an oscillatory
state describing waves of biomass moving up the size spectrum. When averaged
over time, these oscillations average out to a power-law abundance. We intend to
publish these results in a future paper.

Our model is a trait-based rather than a species-based, which means that, 
rather than taking a finite set of species, it uses a continuum of species 
distinguished by a continuous trait variable, in our case the maximum size 
of a cell of that species. All analytical results in this paper very much 
rely on the existence of this continuum of species, which is clearly an 
idealisation of a real aquatic ecosystem that can only contain a finite 
set of species. In that case the community abundance can
never be an exact power law, and therefore also the resulting allometric scaling
can not be exact. One may wonder whether the qualitative results of this paper
will continue to apply. To test that, we have made numerical simulations of a
discrete species version of the model elsewhere \cite{law:2016}, including also
density dependence in the predation to stabilise the steady state. The results
there show that while with only a small number of species the community is very
far from being described by a Sheldon spectrum, with a larger number of species 
the community moves closer to a Sheldon spectrum and also
the stability of the steady state is increased. This is a very interesting result,
because it contradicts a common belief, originating from work of May 
\cite{may:1972}, that the stability of a community decreases
as species richness and connectance increase.

Our model works with a single trait variable. It ignores all
other characteristics that distinguish different species, except whether it is
an autotroph or a heterotroph. Clearly the model could be made more realistic by
also distinguishing between different functional types, for example between
diatoms and dinoflagelates. Also, it would be easy to include mixotrophs without
changing the conclusions of our model. However we wanted our model to be the
simplest conceptual model that clarifies how coexistence on a Sheldon spectrum
emerges. For the same reason, we included only a single resource described
by a very simple equation, whereas in reality the resource dynamics are
complicated and very seasonal. 

One question that we have not addressed in this paper is the reason for the
observed allometric scaling of the phytoplankton growth rate and the predation
rate. As these are at the basis of our derivation of coexistence and the Sheldon
spectrum, finding an explanation for them would be very interesting. The
allometry of the phytoplankton growth rate may possibly be imposed by physical
constraints. The predation kernel, on the other hand, combines two ingredients:
a preferred prey size and a foraging term. Although there may also be physical
constraints for the latter, both ingredients are to a great extend behavioural
---hence subject to evolution. Take the preference for a prey size, for
instance. It is hard to believe that if the abundance of the preferred prey is
seriously depleted the predator will not adapt its consumption habits to keep a
sufficient food supply. We believe that, instead of an input, the predation
kernel should be an emergent feature, consequence of an underlying evolutionary
principle that guides efficient predation habits. We have to leave this
interesting question for the future.

\begin{acknowledgements}
JAC was supported by the Spanish mobility grant PRX12/00124 and project
FIS2015-64349-P (MINECO/FEDER, UE). GWD and RL were supported by EU Grant
634495 MINOUW H2020-SFS-2014-2015.
\end{acknowledgements}

\appendix

\section{Check of scale-invariance of steady-state equations}

In this appendix we will verify our claim that the pair of functions
\begin{align}
\phy_\lambda(w,\ws)=\lambda^{\gamma+1}\phy(\lambda w,\lambda\ws),\qquad
\zoo_\lambda(w,\ws)=\lambda^{\gamma+1}\zoo(\lambda w,\lambda\ws),
\end{align}
where
\begin{equation}
\gamma=1+\nu+\xi,
\label{gamma}
\end{equation}
solve steady state equations provided the original functions $\phy(w,\ws)$,
$\zoo(w,\ws)$ do for the same \resource value $\hat\N$.

The steady state equations, obtained by setting the time derivative to zero in
the dynamical equations \eqref{eq:MFdiv} for $\phy(w,\ws)$, \eqref{eq:pred} for
$\zoo(w,\ws)$, and \eqref{eq:resourcehat} for $\hat{\N}$ are
\begin{align}
\frac{\partial}{\partial w}\big[\gp(w,\ws)\phy(w,\ws)\big] =
&2\int_0^{\ws}\splt(w|w')\dup(w',\ws)\phy(w',\ws)\,dw' \nonumber\\
&-\dup(w,\ws)\phy(w,\ws)-\death(w,\ws)\phy(w,\ws).
\label{psse}\\
\frac{\partial}{\partial w}\big[\gz(w,\ws)\zoo(w,\ws)\big] =
&2\int_0^{\infty}\splt(w|w')\dup_\zoo(w',\ws)\zoo(w',\ws)\,dw' 
\nonumber\\
&-\dup_\zoo(w,\ws)\zoo(w,\ws)-\death(w,\ws)\zoo(w,\ws),
\label{zsse}\\
\hat\varrho(\hat\N)=&\hat\sigma(\hat\N,\phy).
\label{nsse}
\end{align}
To check that $\phy_\lambda(w,\ws)$, $\zoo_\lambda(w,\ws)$ solve these
equations we simply substitute them. Let us start with \eqref{psse} and
consider each term individually. The left-hand side gives
\begin{equation}
\begin{split}
\frac{\partial}{\partial w}\left[\gp(w,\ws)\phy_\lambda(w,\ws)\right]
&=\frac{\partial}{\partial w}\left[\gp(w,\ws)\lambda^{\gamma+1}
\phy(\lambda w,\lambda \ws)\right]\\
&=\lambda\frac{\partial}{\partial(\lambda w)}
\left[\lambda^{\xi-1}\gp(\lambda w,\lambda \ws)\lambda^{\gamma+1}
\phy(\lambda w,\lambda \ws)\right] \\
&=\lambda^{\gamma+1+\xi}\frac{\partial}{\partial(\lambda w)}
\left[\gp(\lambda w,\lambda \ws)\phy(\lambda w,\lambda \ws)\right],
\end{split}
\end{equation}
where we used the scaling property \eqref{eq:invarianceg} of the 
growth rate $\gp(w,\ws)$.
The first term on the right-hand side of \eqref{psse} gives
\begin{equation}
\begin{split}
2\int_0^{\ws}&\splt(w|w')\dup(w',\ws)\phy_\lambda(w',\ws)\,dw'\\
&=2\int_0^{\lambda\ws}\lambda\splt(\lambda w|\lambda w')\lambda^\xi
\dup(\lambda w',\lambda\ws)\lambda^{\gamma+1}\phy(\lambda w',\lambda\ws)
\,\lambda^{-1}d(\lambda w')\\
&=\lambda^{\gamma+1+\xi}
2\int_0^{\lambda\ws}\splt(\lambda w|\lambda w')
\dup(\lambda w',\lambda\ws)\phy(\lambda w',\lambda\ws)
\,d(\lambda w'),
\end{split}
\end{equation}
where we used the scaling properties \eqref{eq:invarianceq} and 
\eqref{eq:invariancek}.
The second term gives
\begin{equation}
\begin{split}
-\dup(w,\ws)\phy_\lambda(w,\ws)
&=-\lambda^\xi \dup(\lambda w,\lambda\ws)\lambda^{\gamma+1}
\phy(\lambda w,\lambda\ws)\\
&=-\lambda^{\gamma+1+\xi} \dup(\lambda w,\lambda\ws)\phy(\lambda w,\lambda\ws)
\end{split}
\end{equation}
again due to the scaling property \eqref{eq:invariancek} of the division rate.
Finally, using the expression \eqref{eq:mupred} for the mortality rate, 
the last term gives
\begin{equation}
\begin{split}
-\death(w,\ws)\phy_\lambda(w,\ws)
&=-\left[\int_0^\infty\pred(w',w)\int_0^\infty\zoo_\lambda(w',\ws')\,d\ws'\,dw'+
\ws^{-\xi}\sdeath_b(w/\ws)\right]\phy_\lambda(w,\ws)\\
&=-\Big[\int_0^\infty\lambda^{-\nu}\pred(\lambda w',\lambda w)
\int_0^\infty\lambda^{\gamma+1}\zoo(\lambda w',\lambda\ws')\,
\lambda^{-1}d(\lambda \ws')\,\lambda^{-1}d(\lambda w')\\
&\qquad\qquad+
\lambda^\xi(\lambda\ws)^{-\xi}\sdeath_b(\lambda w/\lambda\ws)\Big]
\lambda^{\gamma+1}\phy(\lambda w,\lambda\ws)\\
&=-\lambda^{\gamma+1+\xi}\death(\lambda w,\lambda \ws)
\phy_\lambda(\lambda w,\lambda\ws).
\end{split}
\end{equation}
We substituted both $p_\lambda$ and $z_\lambda$, used the scaling property
of the predation rate $S(w,w')$ that follows from Eq.\eqref{eq:predscaling}
and the relation \eqref{gamma} between the exponents.

Putting these four terms back together, we see that the resulting equation is the 
same as the original equation evaluated at scaled weights $\lambda w$ and 
$\lambda\ws$, up to an overall factor of $\lambda^{\gamma+1+\xi}$. 
Given that the original equation holds for all weights $w$ and $\ws$, this shows 
that the transformed equation is equivalent to the original one, establishing its
scale invariance.

In the equation \eqref{zsse} for the zooplankton abundance the left-hand side
involves the zooplankton growth rate given in Eq.~\eqref{eq:h} and we first 
determine its behaviour when $\phy_{\lambda}$ and $\zoo_{\lambda}$ replace the
original functions:
\begin{equation}
\begin{split}
\gz(w,\ws)&=\int_0^\infty\pred(w,w')\epsilon w'
\left[\int_0^\infty(\phy_\lambda(w',\ws')+\zoo_\lambda(w',\ws')) d\ws'\right]dw'
-b\ws^{1-\xi}\left(\frac{w}{\ws}\right)^{\beta}\\
&=\int_0^\infty\lambda^{-\nu}\pred(\lambda w,\lambda w')\epsilon 
\lambda^{-1}(\lambda w')
\left[\int_0^\infty\lambda^{\gamma+1}(\phy(\lambda w',\lambda \ws')
+\zoo(\lambda w',\lambda\ws')) \lambda^{-1}d(\lambda\ws')\right]
\lambda^{-1}d(\lambda w')\\
&\qquad\qquad-b\lambda^{-1+\xi}(\lambda\ws)^{1-\xi}
\left(\frac{\lambda w}{\lambda \ws}\right)^{\beta}\\
&=\lambda^{\xi-1}\gz(\lambda w,\lambda \ws).
\end{split}
\end{equation}
Thus under the scale transformation the left-hand side of the zooplankton
equation \eqref{zsse} becomes
\begin{equation}
\begin{split}
\frac{\partial}{\partial w}\left[\lambda^{\xi-1}\gz(\lambda w,\lambda\ws)
\lambda^{\gamma+1}\zoo(\lambda w,\lambda \ws)\right]
&=\lambda^{\gamma+1+\xi}\frac{\partial}{\partial(\lambda w)}
\left[\gz(\lambda w,\lambda \ws)\zoo(\lambda w,\lambda \ws)\right].
\end{split}
\end{equation}
The terms on the right-hand side transform just like those in the phytoplantkon
equation. So again we find that the transformed equation is the same as the
original equation at rescaled weights up to an overall factor of
$\lambda^{\gamma+1+\xi}$.

Finally, in the renormalised resource equation \eqref{nsse} the left-hand side
does not depend on weights or plankton abundances, so is invariant under the
scale transformation. The right hand side transforms to
\begin{equation}
\begin{split}
\hat\sigma(\hat\N,\phy_\lambda)&=\lim \dfrac{\frac{\hat{a}(\hat\N)}{\theta}
\int_{w_{min}}^{w_{max}}\ws^{1-\alpha-\xi}\int_0^{\ws} w^\alpha
\phy_\lambda(w,\ws)dw d\ws}
{\int_{w_{min}}^{w_{max}}\ws^{1-\xi-\gamma}d\ws}\\
&=\lim \dfrac{\frac{\hat{a}(\hat\N)}{\theta}
\int_{\lambda w_{min}}^{\lambda w_{max}}\lambda^{-1+\alpha+\xi}
(\lambda\ws)^{1-\alpha-\xi}\int_0^{\lambda\ws} \lambda^{-\alpha}(\lambda w)^\alpha
\lambda^{\gamma+1}\phy(\lambda w,\lambda \ws)
\lambda^{-1}d(\lambda w)\lambda^{-1} d(\lambda\ws)}
{\int_{\lambda w_{min}}^{\lambda w_{max}}\lambda^{-1+\xi+\gamma}
(\lambda\ws)^{1-\xi-\gamma}\lambda^{-1}d(\lambda\ws)}\\
&=\hat\sigma(\hat\N,\phy)
\end{split}
\end{equation}
and thus is also invariant, meaning the entire equation is invariant. This
completes the proof that all the steady-state equations are scale-invariant.

\bibliographystyle{spmpsci}      
\bibliography{ecology}   

\begin{thebibliography}{10}
\providecommand{\url}[1]{{#1}}
\providecommand{\urlprefix}{URL }
\expandafter\ifx\csname urlstyle\endcsname\relax
  \providecommand{\doi}[1]{DOI~\discretionary{}{}{}#1}\else
  \providecommand{\doi}{DOI~\discretionary{}{}{}\begingroup
  \urlstyle{rm}\Url}\fi

\bibitem{andersen:2006}
Andersen, K.H., Beyer, J.E.: Asymptotic size determines species abundance in
  the marine size spectrum.
\newblock Am. Nat. \textbf{168}, 54--61 (2006)

\bibitem{armstrong:1999}
Armstrong, R.A.: Stable model structures for representing biogeochemical
  diversity and size spectra in plankton communities.
\newblock J. Plankton Res. \textbf{21}(3), 445--464 (1999)

\bibitem{baird:2007}
Baird, M.E., Suthers, I.M.: A size-resolved pelagic ecosystem model.
\newblock Ecol. Model. \textbf{203}(3), 185--203 (2007)

\bibitem{banas:2011}
Banas, N.S.: Adding complex trophic interactions to a size-spectral plankton
  model: Emergent diversity patterns and limits on predictability.
\newblock Ecol, Model. \textbf{222}(15), 2663--2675 (2011)

\bibitem{benoit:2004}
Beno{\^i}t, E., Rochet, M.J.: A continuous model of biomass size spectra
  governed by predation and the effects of fishing.
\newblock J.\ Theor.\ Biol. \textbf{226}, 9--21 (2004)

\bibitem{camacho:2001}
Camacho, J., Sol{\'e}, R.V.: Scaling in ecological size spectra.
\newblock Eurphys.\ Lett. \textbf{55}, 774--780 (2001)

\bibitem{capitan:2010}
Capit{\'a}n, J.A., Delius, G.W.: Scale-invariant model of marine population
  dynamics.
\newblock Physical Review \textbf{81}, 061,901 (2010)

\bibitem{datta:2010}
Datta, S., Delius, G.W., Law, R.: A jump-growth model for predator-prey
  dynamics: Derivation and application to marine ecosystems.
\newblock Bull.\ Math.\ Biol. \textbf{72}, 1361--1382 (2010)

\bibitem{datta:2011}
Datta, S., Delius, G.W., Law, R., Plank, M.J.: A stability analysis of the
  power-law steady state of marine size spectra.
\newblock Journal of Mathematical Biology \textbf{63}, 779--799 (2011)

\bibitem{delong:2012}
DeLong, J.P., Vasseur, D.A.: Size-density scaling in protists and the links
  between consumer-resource interaction parameters.
\newblock J. Anim. Ecol. \textbf{81}, 1193--1201 (2012)

\bibitem{diekmann:1983}
Diekmann, O., Lauwerier, H.A., Aldenberg, T., Metz, J.A.J.: Growth, fission and
  the stable size distribution.
\newblock J.\ Math.\ Biol. \textbf{18}, 135--148 (1983)

\bibitem{fox:2013}
Fox, J.W.: The intermediate disturbance hypothesis should be abandoned.
\newblock Trends Ecol. Evol. \textbf{28}, 86--92 (2013)

\bibitem{fredrickson:1967}
Fredrickson, A.G., Ramkrishna, D., Tsuchyia, H.M.: Statistics and dynamics of
  procaryotic cell populations.
\newblock Math.\ Biosci. \textbf{1}, 327--374 (1967)

\bibitem{friedlander:2008}
Friedlander, T., Brenner, N.: Cellular properties and population asymptotics in
  the population balance equation.
\newblock Phys.\ Rev.\ Lett. \textbf{101}, 018,104 (2008)

\bibitem{gaedke:1992}
Gaedke, U.: The {{Size Distribution}} of {{Plankton Biomass}} in a {{Large
  Lake}} and its {{Seasonal Variability}}.
\newblock Limnol. Oceanogr. \textbf{37}(6), 1202--1220 (1992)

\bibitem{gin:1998}
Gin, K.Y.H., Guo, J., Cheong, H.F.: A size-based ecosystem model for pelagic
  waters.
\newblock Ecol. Model. \textbf{112}(1), 53--72 (1998)

\bibitem{giometto:2013}
Giometto, A., Altermatt, F., Carrara, F., Maritan, A., Rinaldo, A.: Scaling
  body size fluctuations.
\newblock Proc.\ Nat.\ Acad.\ Sci.\ USA \textbf{110}, 4646--4650 (2013)

\bibitem{guiet:2016}
Guiet, J., Poggiale, J.C., Maury, O.: Modelling the community size-spectrum:
  recent developments and new directions.
\newblock Ecol. Model. \textbf{337}, 4--14 (2016)

\bibitem{hartvig:2011}
Hartvig, M., Andersen, K.H., Beyer, J.E.: Food web framework for
  size-structured populations.
\newblock J.\ Theor.\ Biol. \textbf{272}, 113--122 (2011)

\bibitem{heijmans:1984}
Heijmans, H.J.A.M.: On the stable size distribution of populations reproducing
  by fission into two unequal parts.
\newblock Math.\ Biosci. \textbf{72}, 19--50 (1984)

\bibitem{henson:2003}
Henson, M.A.: Dynamic modeling of microbial cell populations.
\newblock Curr. Opin. Biotechnol. \textbf{14}, 460--467 (2003)

\bibitem{herbert:1956}
Herbert, D., Elsworth, R., Telling, R.C.: The continuous culture of bacteria; a
  theoretical and experimental study.
\newblock J. Gen. Microbiol. \textbf{14}, 601--622 (1956)

\bibitem{hutchinson:1961}
Hutchinson, G.: The paradox of the plankton.
\newblock Am.\ Nat. \textbf{95}, 137--145 (1961)

\bibitem{kempes:2011}
Kempes, C.P., Dutkiewicz, S., Follows, M.J.: Growth, metabolic partitioning,
  and the size of microorganisms.
\newblock Proc.\ Nat.\ Acad.\ Sci.\ USA \textbf{109}, 495--500 (2011)

\bibitem{law:2016}
Law, R., Cuesta, J.A., Delius, G.: Plankton: the paradox and the power law
  (2016).
\newblock Submitted for publication

\bibitem{leibold:1996}
Leibold, M.A.: Graphical model of keystone predators in food webs: trophic
  regulation of abundance, incidence, and diversity patterns in communities.
\newblock Am. Nat. \textbf{147}, 784--812 (1996)

\bibitem{maranon:2013}
Mara{\~n\'o}n, E., Cerme{\~n}o, P., L{\'o}pez-Sandoval, D.C.,
  Rodr{\'{\i}}guez-Ramos, T., Sobrino, C., Huete-Ortega, M., Blanco, J.M.,
  Rodr{\'{\i}}guez, J.: Unimodal size scaling of phytoplankton growth and the
  size dependence of nutrient uptake and use.
\newblock Ecol. Lett. \textbf{16}, 371--379 (2013)

\bibitem{may:1972}
May, R.M.: Will a large complex system be stable?
\newblock Nature \textbf{238}, 413--414 (1972)

\bibitem{moloney:1991}
Moloney, C.L., Field, J.G.: The size-based dynamics of plankton food webs. i. a
  simulation model of carbon and nitrogen flows.
\newblock J. Plankton Res. \textbf{13}(5), 1003--1038 (1991)

\bibitem{platt:1977}
Platt, T., Denman, K.: Organisation in the pelagic ecosystem.
\newblock Helgol. Mar. Res. \textbf{30}, 575--581 (1977)

\bibitem{poulin:2010}
Poulin, F.J., Franks, P.J.S.: Size-structured planktonic ecosystems:
  constraints, controls and assembly instructions.
\newblock J. Plankton Res. \textbf{32}(8), 1121--1130 (2010)

\bibitem{powell:1964}
Powell, E.O.: A note on koch \& schaechter’s hypothesis about growth and
  fission of bacteria.
\newblock J. Gen. Microbiol. \textbf{37}, 231--249 (1964)

\bibitem{quinones:2003}
Qui{\~n}ones, R.A., Platt, T., Rodr\'{\i}guez, J.: Patterns of biomass-size
  spectra from oligotrophic waters of the northwest atlantic.
\newblock Prog. Oceanogr. \textbf{57}, 405--427 (2003)

\bibitem{roy:2007}
Roy, S., Chattopadhyay, J.: Towards a resolution of `the paradox of the
  plankton': a brief overview of the proposed mechanisms.
\newblock Ecol. Compl. \textbf{4}, 26--33 (2007)

\bibitem{san_martin:2006}
San~Martin, E., Harris, R.P., Irigoien, X.: Latitudinal variation in plankton
  size spectra in the atlantic ocean.
\newblock Deep-Sea Res. Pt. {II} \textbf{53}(14), 1560--1572 (2006)

\bibitem{sheldon:1972b}
Sheldon, R.W., Kerr, S.: The population density of monsters in {Loch Ness}.
\newblock Limnol. Oceanogr. \textbf{17}, 796--799 (1972)

\bibitem{sheldon:1972a}
Sheldon, R.W., Prakash, A., Sutcliffe, W.H.: The size distribution of particles
  in the ocean.
\newblock Limnol. Oceanogr. \textbf{17}, 327--340 (1972)

\bibitem{silvert:1978}
Silvert, W., Platt, T.: Energy flux in the pelagic ecosystem: A time-dependent
  equation.
\newblock Limnol. Oceanogr. \textbf{23}, 813--816 (1978)

\bibitem{silvert:1980}
Silvert, W., Platt, T.: Dynamic energy-flow model of the particle size
  distribution in pelagic ecosystems.
\newblock In: W.C. Kerfoot (ed.) Evolution and ecology of zooplankton
  communities, pp. 754--763. New England University Press, Hanover, New
  Hampshire (1980)

\bibitem{sprules:2016}
Sprules, W.G., Barth, L.E., Giacomini, H.: Surfing the biomass size spectrum:
  some remarks on history, theory, and application.
\newblock Can. J. Fish. Aquat. Sci. \textbf{73}(4), 477--495 (2016)

\bibitem{stock:2008}
Stock, C.A., Powell, T.M., Levin, S.A.: Bottom-up and top-down forcing in a
  simple size-structured plankton dynamics model.
\newblock J. Marine Syst. \textbf{74}(1), 134--152 (2008)

\bibitem{thingstad:1997}
Thingstad, T.F., Lignell, R.: Theoretical models for the control of bacterial
  growth rate, abundance, diversity and carbon demand.
\newblock Aquat. Microb. Ecol. \textbf{13}, 19--27 (1997)

\bibitem{tyson:1986}
Tyson, J.J., Diekmann, O.: Sloppy size controlo of the cell division cycle.
\newblock J.\ Theor.\ Biol. \textbf{118}, 405--426 (1986)

\bibitem{vonbertalanffy:1957}
{von Bertalanffy}, L.: Quantitative laws in metabolism and growth.
\newblock Quart.\ Rev.\ Biol. \textbf{32}, 217--231 (1957)

\bibitem{vage:2014}
V\r{a}ge, S., Storesund, J.E., Giske, J., Thingstad, T.F.: Optimal defense
  strategies in an idealized microbial food web under trade-off between
  competition and defense.
\newblock PLoS ONE \textbf{9}, e101,415 (2014)

\bibitem{ward:2014}
Ward, B.A., Dutkiewicz, S., Follows, M.J.: Modelling spatial and temporal
  patterns in size-structured marine plankton communities: top–down and
  bottom–up controls.
\newblock J. Plankton Res. \textbf{36}(1), 31--47 (2014)

\bibitem{winter:2010}
Winter, C., Bouvier, T., Weinbaue, M.G., Thingstad, T.F.: Trade-offs between
  competition and defense specialists among unicellular planktonic organisms:
  the ``killing the winner'' hypothesis revisited.
\newblock Microb. Molec. Biol. Rev. \textbf{74}, 42--57 (2010)

\bibitem{wirtz:2012}
Wirtz, K.W.: Who is eating whom? {M}orphology and feeding type determine the
  size relation between planktonic predators and their ideal prey.
\newblock Mar. Ecol.-Prog. Ser. \textbf{445}, 1--12 (2012)

\end{thebibliography}

%
%

\end{document}